\documentclass[aps,pra,a4paper, twocolumn, 10pt, superscriptaddress]{revtex4-1}

\pdfoutput=1
\usepackage{empheq}
\usepackage{dcolumn}
\usepackage{amsmath,amssymb}
\usepackage{bm}
\usepackage{color}
\usepackage{overpic}
\usepackage{latexsym}
\usepackage{epstopdf}
\usepackage{color}
\usepackage[english]{babel}
\makeatletter\AtBeginDocument{\let\@elt\relax}\makeatother

\usepackage{graphicx}
\usepackage{psfrag} 
\usepackage{epsf} 
\usepackage{amsfonts}
\usepackage{bm}
\usepackage{appendix}
\usepackage{scalerel}

\definecolor{mygrey}{gray}{0.35}
\definecolor{myblue}{rgb}{0.2,0.2,0.8}
\definecolor{myzard}{cmyk}{0,0,0.05,0}
\definecolor{mywhite}{rgb}{1,1,1}
\definecolor{myred}{rgb}{1,0.,0.3}

\usepackage{hyperref}
\def\beq{\begin{equation}}
\def\eeq{\end{equation}}
\def\ba{\begin{align}}
\def\enda{\end{align}}
\def\bi{\begin{itemize}}
\def\ei{\end{itemize}}

\newcommand{\ket}[1]{|#1\rangle}
\newcommand{\bra}[1]{\langle #1|}
\newcommand{\braket}[2]{\langle #1|#2\rangle}

\usepackage[normalem]{ulem}

\definecolor{forestgreen(web)}{rgb}{0.13, 0.55, 0.13}

\begin{document}

\title{Strongly bound fermion pairs on a ring: a composite-boson approach}

\author{E. Cuestas}
\affiliation{Instituto de F\'{i}sica Enrique Gaviola, CONICET and Universidad 
Nacional de C\'{o}rdoba,
Ciudad Universitaria, X5016LAE, C\'{o}rdoba, Argentina}
\author{C. Cormick}
\affiliation{Instituto de F\'{i}sica Enrique Gaviola, CONICET and Universidad 
Nacional de C\'{o}rdoba,
Ciudad Universitaria, X5016LAE, C\'{o}rdoba, Argentina}

% \date{\today}
\begin{abstract}
Particles made of two fermions can in many cases be treated as elementary bosons, but the conditions for this treatment to be valid are nontrivial. The so-called ``coboson formalism'' is a powerful tool to tackle compositeness effects relevant for instance for exciton physics and ultracold atomic dimers. A key element of this theory is an ansatz for the ground state of $N$ pairs, built from the single-pair ground state combined with the exclusion principle. We show that this ansatz can fail in one-dimensional systems which fulfill the conditions expected to make the ansatz valid. Nevertheless, we also explain how coboson theory can recover the correct ground state. Thus, our work highlights limitations and strengths of the formalism and leads to a better treatment of composite bosons.
\end{abstract}

\maketitle

\section{Introduction}

The idea that composite particles made of an even number of fermions can behave as elementary bosons is expected from the fact that they have integer total spin, or, equivalently, that the quantum state must be symmetric under exchange of two such composite particles. These observations have been related with experimental results such as superfluidity, superconductivity, or Bose-Einstein condensation of bosonic atoms \cite{Annett_book}. However, a careful treatment of compositeness effects has been elusive until the last decades.

Coboson theory (see Refs. \cite{Combescot_2008, Combescot2015excitons}) has its origin in the study of the conditions for approximately bosonic behavior in excitonic systems~\cite{Combescot_2001}. The discovery that compositeness effects could be relevant in regimes for which they were assumed negligible sparked the motivation for a more accurate description of the associated phenomena \cite{Combescot_2003}. Further developments of the theory included its application to Bose-Einstein condensates \cite{Combescot_2008b}, Cooper pairs \cite{Combescot_2011_Cooper, Combescot_2013_BCS} and the extension to non-zero temperatures \cite{Combescot_2011}. In parallel, a quantum information approach provided a deeper insight in the behavior of composite bosons showing a connection between the bosonic character and the entanglement of the constituents \cite{Law_2005, chudzicki_2010, tichy_2012b, combescot_2011b, ramanathan_2011}.

In this work we investigate the validity of the so-called coboson ansatz for the ground state, which plays an important role in coboson theory. This ansatz approximates the ground state of a many-body system by the repeated application of the operator that, when acting on the vacuum, creates a single pair in its ground state. This procedure was shown to be valid for excitons in Ref. \cite{Keldysh_1968}, and is much older and more widely used than coboson formalism. We refer to such approximation as coboson ansatz because it has crossed the borders of exciton physics to other areas in which composite bosons are relevant. This includes the examples of Wigner molecules~\cite{Cuestas_2020} and specially ultracold atomic dimers \cite{Combescot_2016, Shiau_2019, Bouvrie_2019}. In particular, \cite{Bouvrie_2017} used the coboson ansatz to explore the BCS-BEC crossover that can be experimentally realized using Feshbach resonances \cite{Regal_2003, Chin_2004}.

Conceptually, the ansatz is related with the intuition that the ground state of a system of composite bosons should be similar to a product state where each composite particle occupies the same orbital. It can be shown that this is the canonical-ensemble counterpart of the BCS ansatz \cite{Combescot_2013_BCS}.  We note that coboson theory can be applied even when the ansatz is not valid, as illustrated by the treatment of Cooper pairs \cite{Combescot_2011_Cooper, Combescot_2013_BCS, Combescot2015excitons}. However, this approximation of the ground state greatly simplifies the description of many-body systems.

The coboson ansatz is expected to be valid for short-ranged interactions, dilute systems, and highly entangled constituents. A recent article \cite{Cespedes_2019} has shown that the coboson ansatz may break down in a one-dimensional (1D) discrete model, even when it satisfies the above-mentioned conditions. Although these results were not surprising in the light that strongly bound fermion pairs behave as hard-core bosons \cite{Guan_2013, Girardeau_1960}, the consequences of this behavior for the validity of the ansatz in 1D systems have not been studied in detail so far.

Here we present a comparison between the coboson ansatz and the true ground state for a model of two fermion pairs with delta-interactions moving in one dimension with periodic boundary conditions, and we show that the spatial structure and the energy are poorly predicted by the coboson ansatz also for this continuous system. In the limit of infinite attraction, the fidelity between the two states coincides with the one found for the discrete model of \cite{Cespedes_2019}. Also in coincidence with Ref. \cite{Cespedes_2019}, we associate the main difference between the two descriptions with the presence of strong spatial correlations between pairs. Still, by using coboson theory we develop a perturbative calculation that readily recovers the correct results. Thus, although our study stresses the limitations of the ansatz, it also illustrates the power of the coboson formalism to describe composite bosons in the regime of strong interaction. 

The article is organized as follows: in Sec.~\ref{sec:coboson} we provide a brief summary of concepts of coboson theory that are essential for our work. In Sec.~\ref{sec:model} we explain the model we study, and in Sec.~\ref{sec:writing}
we apply the tools of the coboson formalism to the system under consideration. Section~\ref{sec:solution} describes the ground state of the system and compares it with the coboson ansatz. In Sec.~\ref{sec:conclusion} we state our conclusions. Technical details are contained in several Appendices.

%%%%%%%%%%%%%%%%%%%
%%%%%%%%%%%%%%%%%%%
% Sect: Coboson formalism and coboson ansatz
%%%%%%%%%%%%%%%%%%%
%%%%%%%%%%%%%%%%%%%
\section{Coboson formalism and coboson ansatz}
\label{sec:coboson}

We present here some aspects of coboson theory that are essential for our purposes; a very detailed introduction can be found in \cite{Combescot_2008}. We focus on the case of particles composed by two fermions of different kinds ($a$ and $b$), such as different sorts of identical fermions or fermions with different spin in models with no spin flips.

The basic elements of the formalism are the coboson creation operators $B^\dagger_j$ that, when acting on the vacuum $\ket{v}$, create a single-pair eigenstate $\ket{\alpha_j}$ of the Hamiltonian.  In a separable reference basis, they can be written as:
\begin{equation}
 B^\dagger_j = \sum_{q,p} \braket{q,p}{\alpha_j} \, a^\dagger_q b^\dagger_p \,. 
\end{equation}
Here, $a^\dagger_q$ ($b^\dagger_p$) are creation operators for fermions of kind $a$ ($b$), with $p$ and $q$ labelling the states in the separable basis and $\braket{q,p}{\alpha_j}$ the matrix for basis change.

Of special relevance is the coboson creation operator $B_0^\dagger$ which, when acting on $\ket{v}$, creates a single pair in its ground state. This operator is essential for the proposed approximation for the many-body ground state of $N$ pairs:
%%%%%%%%%%%%%%%%%%%
%%%%%%%%%%%%%%%%%%%
\begin{equation}
 \ket{N} = \frac{1}{\sqrt{F_N N!}} (B_0^\dagger)^N \ket{v} \,,
\end{equation}
%%%%%%%%%%%%%%%%%%%
%%%%%%%%%%%%%%%%%%%
\noindent where the normalization constant $F_N$ depends on the properties of the single-pair ground state. In particular, $F_2$ and the ratios $F_N/F_{N-1}$ approach the bosonic value of~$1$ when the single-pair constituents are highly entangled \cite{Law_2005, chudzicki_2010, tichy_2012b}. The state $\ket{N}$, or coboson ansatz, is expected to provide a good description of the full ground state when the system is dilute and the interactions are sufficiently short-ranged \cite{Combescot_2008}. The ground-state energy $E_{\rm GS}^{(N)}$ for $N$ pairs is then approximated by the expectation value $E_{\rm cob}^{(N)}$ of the Hamiltonian in the state $\ket{N}$:
\begin{equation}
 E_{\rm GS}^{(N)} \simeq E_{\rm cob}^{(N)} = \langle N| H |N\rangle   
\end{equation}

Taking the coboson ansatz as starting point, the coboson formalism naturally leads to expressions in the form of expansions in powers of density or number of particles, arising from the commutators of the relevant operators~\cite{Combescot_2008}. In particular, the estimated ground-state energy of $N$ pairs has its main contributions determined by the values for 1 and 2 pairs as:
%%%%%%%%%%%%%%%%%%%
%%%%%%%%%%%%%%%%%%%
\begin{equation}
 E_{\rm cob}^{(N)} \simeq N E_{\rm GS}^{(1)} + \frac{N(N-1)}{2} \frac{F_2 F_{N-2}}{F_N} \Big(E_{\rm cob}^{(2)} - 2 E_{\rm GS}^{(1)}\Big) .
\end{equation}
%%%%%%%%%%%%%%%%%%%
%%%%%%%%%%%%%%%%%%%
\noindent Here, both $F_2$ and the ratio $F_{N-2}/F_N$ are close to 1 in the limit when the pairs approach perfecly bosonic behaviour \cite{Combescot_2008, chudzicki_2010}. In this case, the above equation can be further approximated by:
\begin{equation}
 E_{\rm cob}^{(N)} \simeq N E_{\rm GS}^{(1)} + \frac{N(N-1)}{2} \Big(E_{\rm cob}^{(2)} - 2 E_{\rm GS}^{(1)}\Big) .
 \label{eq:energy_coboson_N}
\end{equation}
Expressions are also given in Ref.~\cite{Combescot_2008} for the calculation of $E_{\rm cob}^{(2)}$ and $F_N$, but we omit them because we will follow a slightly different path.

%%%%%%%%%%%%%%%%%%%
%%%%%%%%%%%%%%%%%%%
% Sect: The model
%%%%%%%%%%%%%%%%%%%
%%%%%%%%%%%%%%%%%%%
\section{The physical model}
\label{sec:model}

We consider fermions of kinds $a$ and $b$, both of mass $m$ and moving in a 1D system of length $L$ with periodic boundary conditions. We assume that the creation and annihilation operators corresponding to different fermion species commute; this choice does not affect the final results. Particles of different kind interact through a delta-type attraction:
%%%%%%%%%%%%%%%%%%%
%%%%%%%%%%%%%%%%%%%
\begin{equation}
 H_{\rm int} = -\gamma \int dx \Psi^\dagger_a (x) \Psi^\dagger_b (x) \Psi_a (x) \Psi_b (x) .
 \label{eq:Hint}
\end{equation}
%%%%%%%%%%%%%%%%%%%
%%%%%%%%%%%%%%%%%%%

Our main goal is to compare the coboson ansatz with the true ground state for two pairs in the regime of strong attraction. We note that, for this particular model, the limit of infinite attraction is the same as the infinitely dilute system, which is when coboson theory is expected to work best. In this strongly attractive limit, fermions of different species make up highly bound pairs that behave like hard-core bosons, approximating a Tonks-Girardeau gas \cite{Tonks_1936, Girardeau_1960}.

Much research has been done on similar problems in ultracold atom settings \cite{Mora_2005, Guan_2009, Chen_2010_bosons, Chen_2010_fermions, Guan_2010, Guan_2013}. In particular, analytical solutions can be found \cite{Gaudin_1967, Yang_1967, Flicker_1967, Takahashi_1971, Gu_1989, Oelkers_2006}, and the ground-state energy for $N$ pairs in the strongly attractive limit is known to be~\cite{Chen_2010_fermions}:
%%%%%%%%%%%%%%%%%%%
%%%%%%%%%%%%%%%%%%%
\begin{equation}
 E_{\rm GS}^{(N)} \simeq \frac{\hbar^2}{m} \left[-N\lambda^2 + \frac{\pi^2}{2L^2} \frac{N(N^2-1)}{6} \left(1+\frac{N}{L\lambda}
 \right) \right] ,
\label{eq:energy_known}
\end{equation}
%%%%%%%%%%%%%%%%%%%
%%%%%%%%%%%%%%%%%%%
\noindent with $\lambda^{-1}$ being the size of each pair (see below for details) and $\lambda L \gg N$ (no thermodynamic limit has been taken). Notice that Eq.~(\ref{eq:energy_known}) presents an important difference with the coboson prediction of Eq.~\eqref{eq:energy_coboson_N} in the dependence of the energy with the number $N$. In what follows we show that the coboson ansatz also gives an incorrect order of magnitude for the energy (subtracting the binding energy).

Writing the wavefunction following the literature \cite{Gaudin_1967, Yang_1967, Flicker_1967, Takahashi_1971, Gu_1989, Oelkers_2006} is cumbersome and makes some calculations rather difficult. Alternative numerical approaches based on the single-particle solutions (in our case, plane waves) converge very slowly in the strongly interacting limit. Here we present a different strategy based on the concepts of Ref. \cite{Combescot_2008}. We study the problem in the coboson basis, i.e. in terms of the eigenstates of one pair of interacting fermions in the ring, but without assuming the coboson ansatz. This approach leads to a very compact description of the approximate ground state that can be readily used for further calculations.

%%%%%%%%%%%%%%%%%%%
%%%%%%%%%%%%%%%%%%%
% Sect: Writing the problem with tools from coboson theory
%%%%%%%%%%%%%%%%%%%
%%%%%%%%%%%%%%%%%%%
\section{Writing the problem with tools from coboson theory}
\label{sec:writing}

\subsection{Single-pair eigenbasis}

To define the coboson basis one must first solve the single-pair problem. In our case, we can separate the relative and center-of-mass degrees of freedom. Moreover, in the strongly attractive regime one can take a restricted basis considering only the ground state of the relative motion and a set of low-lying states of the center of mass. The solutions for the center of mass (CM) have the form:
%%%%%%%%%%%%%%%%%%%
%%%%%%%%%%%%%%%%%%%
\begin{equation}
\psi_{{\scriptscriptstyle CM}}(x_{{\scriptscriptstyle CM}})^{(K)} = \frac{1}{\sqrt{L}} e^{iK\scaleto{x}{4pt}_{{\scriptscriptstyle CM}}}, \quad K = 2\pi k/L  \,,
\end{equation}
%%%%%%%%%%%%%%%%%%%
%%%%%%%%%%%%%%%%%%%
\noindent with $k\in\mathbb{Z}$. The associated motional energy is given by 
%%%%%%%%%%%%%%%%%%%
%%%%%%%%%%%%%%%%%%%
\begin{equation}
 E_K = \frac{\hbar^2 K^2}{4m} \,.
\end{equation}
%%%%%%%%%%%%%%%%%%%
%%%%%%%%%%%%%%%%%%%
\noindent For the relative motion, the energy is:
%%%%%%%%%%%%%%%%%%%
%%%%%%%%%%%%%%%%%%%
\begin{equation}
 E_\gamma = -\frac{\hbar^2\lambda^2}{m} \,,\quad
  \lambda \simeq \frac{m\gamma}{2\hbar^2} ,
\end{equation}
%%%%%%%%%%%%%%%%%%%
%%%%%%%%%%%%%%%%%%%
\noindent up to corrections that are exponentially small for $\lambda L\gg 1$. In the same limit, the wavefunction can be written as:
%%%%%%%%%%%%%%%%%%%
%%%%%%%%%%%%%%%%%%%
\begin{equation}
 \psi_r (x_r) \simeq \sqrt{\lambda}\,\, e^{-\lambda |x_r|} \,.
\end{equation}
%%%%%%%%%%%%%%%%%%%
%%%%%%%%%%%%%%%%%%%
\noindent Here and in the following we neglect terms of the order of $e^{-\lambda L}$ (see Appendix \ref{sec:relative} for more details).

From these solutions we can build the creation operator for a single pair as:
%%%%%%%%%%%%%%%%%%%
%%%%%%%%%%%%%%%%%%%
\begin{multline} 
 B^\dagger_K = \sqrt{\frac{\lambda}{L}} \int dx_{{\scriptscriptstyle CM}}\, dx_r\, e^{iK \scaleto{x}{4pt}_{{\scriptscriptstyle CM}}} e^{-\lambda |x_r|}\\ \Psi^\dagger_a(x_{{\scriptscriptstyle CM}}+x_r/2)\, \Psi^\dagger_b(x_{{\scriptscriptstyle CM}}-x_r/2) ,
\label{eq:coboson_field_basis}
\end{multline}
%%%%%%%%%%%%%%%%%%%
%%%%%%%%%%%%%%%%%%%
\noindent where $\Psi^\dagger_\alpha(x)$ creates a fermion of kind $\alpha$ at position $x$.

\subsection{Two-pair coboson basis}

Because the ground state is translationally invariant, we restrict the two-coboson basis to: $\{ B^\dagger_K B^\dagger_{-K} \ket{v}, K=0,1,2,\ldots K_M\}$ with $K_M$ a cutoff. The goal is to find the coefficients of the ground state (GS) as:
%%%%%%%%%%%%%%%%%%%
%%%%%%%%%%%%%%%%%%%
\begin{equation}
 \ket{\psi_{GS}} \simeq \sum_{K=0}^{K_M} c_K B^\dagger_K B^\dagger_{-K} \ket{v} \,.
 \label{eq:expansion}
\end{equation}
%%%%%%%%%%%%%%%%%%%
%%%%%%%%%%%%%%%%%%%
\noindent The coboson ansatz corresponds to $c_K = 0 ~\forall K \neq 0$.

The different states in the basis are neither orthogonal nor normalized \cite{Combescot_2008}, so we need the overlap matrix $S$:
%%%%%%%%%%%%%%%%%%%
%%%%%%%%%%%%%%%%%%%
\begin{equation}
 S_{PK} = \bra{v} B_{P} B_{-P} B^\dagger_K B^\dagger_{-K} \ket{v} \,.
 \label{eq:Sdef}
\end{equation}
%%%%%%%%%%%%%%%%%%%
%%%%%%%%%%%%%%%%%%%
\noindent The steps in the calculation of $S$ involve replacing the form of the coboson operators, using the (anti)commutation relations, and performing the resulting integrals. The details of the calculation are provided in Appendix \ref{sec:calculations}, and the final result is:
%%%%%%%%%%%%%%%%%%%
%%%%%%%%%%%%%%%%%%%
\begin{widetext}
\begin{equation}
 S_{KP} = \delta_{KP} (1 + \delta_{K0})-\frac{1}{2L\lambda} \frac{10+\frac{K^2+P^2}{(2\lambda)^2} }{ 
% \nonumber \\
  \left[ 1+\frac{(P+K)^2}{(4\lambda)^2} \right]
   \left[ 1+\frac{(P-K)^2}{(4\lambda)^2} \right]
 %  \nonumber \\
  \left[ 1+\frac{P^2}{(2\lambda)^2} \right]
  \left[ 1+\frac{K^2}{(2\lambda)^2} \right]} .
\label{eq:S}
\end{equation}
\end{widetext}
%%%%%%%%%%%%%%%%%%%
%%%%%%%%%%%%%%%%%%%

The norm of the coboson ansatz for $N=2$ is related with the corresponding diagonal term, $S_{00} = 2F_2 = 2-5/(\lambda L)$, showing that the normalization approaches the one for bosonic particles for sufficiently large $\lambda L$. This together with the short-ranged interaction and the increasingly dilute character of the system imply that the usual conditions for the coboson ansatz are fulfilled for $\lambda L\gg 1$. In Appendix \ref{sec:schmidt} we show that the majorization condition proposed in \cite{ramanathan_2011} is also satisfied. 
\bigskip

\subsection{Hamiltonian matrix in the coboson basis}

To evaluate the matrix elements of the Hamiltonian, it is convenient to use \cite{Combescot_2008}:
%%%%%%%%%%%%%%%%%%%
%%%%%%%%%%%%%%%%%%%
\begin{equation}
 H_{KP} = 2 S_{KP} (E_\gamma+ E_P) + \Delta_{KP} \,,
 \label{eq:Hmatrix}
\end{equation}
%%%%%%%%%%%%%%%%%%%
%%%%%%%%%%%%%%%%%%%
\noindent with
%%%%%%%%%%%%%%%%%%%
%%%%%%%%%%%%%%%%%%%
\begin{equation}
 \Delta_{KP} = \bra{v} B_K B_{-K} [ [ H_{\rm int}, B^\dagger_{P} ], B^\dagger_{-P} ] \ket{v} \,. 
 \label{eq:Delta_formula}
\end{equation}
%%%%%%%%%%%%%%%%%%%
%%%%%%%%%%%%%%%%%%%
\noindent The calculation of $\Delta_{KP}$ follows similar steps as for $S$ and the result is:
%%%%%%%%%%%%%%%%%%%
%%%%%%%%%%%%%%%%%%%
\begin{widetext}
\begin{equation}
  \Delta_{KP} = \frac{2 \gamma}{L}
  \left\lbrace \frac{  \left(1 + \frac{P^2+K^2}{(4\lambda)^2} \right) + 2 \left( 1 + \frac{K^2}{(4\lambda)^2} \right) }{ \left( 1 + \frac{K^2}{(2\lambda)^2} \right) \left(1 + \frac{(P+K)^2}{(4\lambda)^2} \right) \left( 1 + \frac{(P-K)^2}{(4\lambda)^2} \right) } 
   - \frac{1}{\left( 1+\frac{(P+K)^2}{(4\lambda)2} \right)^2} - \frac{1}{\left( 1+\frac{(P-K)^2}{(4\lambda)2} \right)^2} \right\rbrace \,.
\end{equation}
\end{widetext}
%%%%%%%%%%%%%%%%%%%
%%%%%%%%%%%%%%%%%%%

Assuming that the coboson ansatz holds, its prediction for the energy gives
%%%%%%%%%%%%%%%%%%%
%%%%%%%%%%%%%%%%%%%
\begin{equation}
 E_{\rm cob}^{(2)} \simeq \frac{H_{00}}{S_{00}} \simeq 2 E_\gamma + \frac{\gamma}{L} ,
 \label{eq:energy_coboson}
\end{equation}
%%%%%%%%%%%%%%%%%%%
%%%%%%%%%%%%%%%%%%%
\noindent in disagreement with the known solution given in Eq.~\eqref{eq:energy_known} by a very large term, of order $\gamma/L$. Apart from the binding energy $2E_\gamma$, the solution in Eq.~\eqref{eq:energy_known} evaluated for $N=2$ only has a contribution of order $\hbar^2\pi^2/(mL^2)$. Therefore, the coboson ansatz provides a very poor approximation of the ground-state energy for this problem. Furthermore, according to the expansion of Eq.~\eqref{eq:energy_coboson_N} this immediately means that the energy estimation for $N$ pairs obtained from the coboson ansatz for arbitrary $N$ with $\lambda L\gg N$ is approximately given by:
%%%%%%%%%%%%%%%%%%%
%%%%%%%%%%%%%%%%%%%
\begin{equation}
 E_{\rm cob}^{(N)} \simeq N E_\gamma + \frac{N(N-1)}{2} \frac{\gamma}{L} ,
\end{equation}
%%%%%%%%%%%%%%%%%%%
%%%%%%%%%%%%%%%%%%%
\noindent which clearly differs from the well-known result of Eq.~\eqref{eq:energy_known}, both in the dependence with $N$ and with the system parameters.

\subsection{Quantities of interest for the characterization of the ground state}

Before calculating the ground state, we mention some quantities of interest other than the energy. In terms of the $c_K$, one can compute the fidelity between the numerical ground state and the coboson ansatz as:
%%%%%%%%%%%%%%%%%%%
%%%%%%%%%%%%%%%%%%%
\begin{equation}
 \mathcal{F} = \frac{|\bra{v} B_0^2\ket{\psi_{{\scriptscriptstyle GS}}}|^2}{2\braket{\psi_{{\scriptscriptstyle GS}}}{\psi_{{\scriptscriptstyle GS}}}} = \frac{|(S c)_0|^2}{2 c\,^t S c} ,
\end{equation}
%%%%%%%%%%%%%%%%%%%
%%%%%%%%%%%%%%%%%%%
\noindent where the $t$-superscript indicates the transpose and $c$ is a vector containing the coefficients of Eq.~(\ref{eq:expansion}).

The results of Ref.~\cite{Cespedes_2019} relate the failure of the coboson ansatz with the presence of long-ranged spatial correlations. We then calculate the conditional probability $P(x|0)$ of finding a particle of kind $a$ at position $x$ given that another particle of the same kind was found at position $0$. This conditional probability is given by the squared norm of the state obtained when applying the operators $\Psi_a(x) \Psi_a(0)$ on the ground state:
%%%%%%%%%%%%%%%%%%%
%%%%%%%%%%%%%%%%%%%
\begin{multline}
 P(x|0) \propto \sum_{K,K'} c_K c_{K'}\\ \bra{v} B_K B_{-K} \Psi_a^\dagger (0) \Psi_a^\dagger(x) \Psi_a(x) \Psi_a (0) B^\dagger_{K'} B^\dagger_{-K'} \ket{v} \,.
 \label{eq:cond_prob_def}
\end{multline}
%%%%%%%%%%%%%%%%%%%
%%%%%%%%%%%%%%%%%%%
\noindent Note that the fermionic character of the particles guarantees $P(0|0)=0$. 

Following similar steps as in the previous calculations, we can also find: % Eq.~\eqref{eq:cond_prob_def}
%%%%%%%%%%%%%%%%%%%
%%%%%%%%%%%%%%%%%%%
\begin{widetext}
\begin{align}
 P(x|0) \propto & \sum_{K,K'} c_K c_{K'} \left\lbrace\rule{0cm}{1cm}\right. 
 \left[\frac{\cos[ (K-K') x]}{ \left( 1+\frac{(K-K')^2}{ (4\lambda)^2} \right)^2 } 
 - e^{-2\lambda|x|} \, \frac{ \cos \left[ \frac{K+K'}{2} \, x \right] }{ \left( 1+ \frac{(K-K')^2}{ (4 \lambda)^2 } \right)^2 }   
{\left( \cos \left[ \frac{K-K'}{4} \, x \right] + \frac{ \sin \left[ \frac{K-K'}{4} \, |x| \right]} {\frac{K-K'}{4 \lambda}} \right)\!}^2 \, \right] \nonumber\\
& 
 + \text{same with } K' \to -K' 
 \vphantom{ \frac{\cos[ (K-K') x]}{ \left( 1+\frac{(K-K')^2}{ (4\lambda)^2} \right)^2} } \left.\rule{0cm}{1cm}\right\rbrace .
 %\label{eq:cond_prob}
\end{align}
\end{widetext}
%%%%%%%%%%%%%%%%%%%
%%%%%%%%%%%%%%%%%%%
\noindent In this expression, for the terms where a vanishing denominator is found the corresponding limit must be taken. The overall proportionality constant can be obtained by imposing that the integral of $P(x|0)$ over $x$ equals 1. Due to the fermionic character of the constituents the result must vanish for $x=0$ regardless of the $c_K$, which is guaranteed by the contribution of the exponentially decaying term. 

In particular, for the coboson ansatz we obtain:
%%%%%%%%%%%%%%%%%%%
%%%%%%%%%%%%%%%%%%%
\begin{equation}
 P(x|0) = \frac{1-e^{-2\lambda |x|}}{L[1-1/(\lambda L)]} \,.
 \label{eq:prob_cob}
\end{equation}
%%%%%%%%%%%%%%%%%%%
%%%%%%%%%%%%%%%%%%%

%%%%%%%%%%%%%%%%%%%
%%%%%%%%%%%%%%%%%%%
% Sect: Ground-state solution in the coboson basis.
%%%%%%%%%%%%%%%%%%%
%%%%%%%%%%%%%%%%%%%
\section{Ground-state solution in the coboson basis} 
\label{sec:solution}

\subsection{Analytical treatment in the limit of very strong attraction}

For very strong attraction, one can find an approximate solution for the ground state using an expansion in $\epsilon=1/(L\lambda)$ within the generalized eigenvalue equation $H c = E S c$. For the overlap matrix the expansion gives:
%%%%%%%%%%%%%%%%%%%
%%%%%%%%%%%%%%%%%%%
\begin{equation}
 S_{KP} \simeq \delta_{KP} (1+\delta_{K0}) -5\epsilon \,.
\label{eq:S_Taylor}
\end{equation}
%%%%%%%%%%%%%%%%%%%
%%%%%%%%%%%%%%%%%%%
\noindent As a consequence of the different spatial scales for center of mass and relative motion, the first order does not depend on $K,P$. For the Hamiltonian, one finds:
%%%%%%%%%%%%%%%%%%%
%%%%%%%%%%%%%%%%%%%
\begin{equation}
 H_{KP} \simeq 2E_\gamma \delta_{KP} + \frac{2\gamma}{L} + \delta_{KP} \frac{\hbar^2 P^2}{2m} - \frac{11 \epsilon \hbar^2}{4 m} (P^2+K^2) \,.
 \label{eq:H_Taylor}
\end{equation}
%%%%%%%%%%%%%%%%%%%
%%%%%%%%%%%%%%%%%%%
\noindent Since the binding energy $2 E_\gamma$ is the same for all the states in our description, we can look for the energy difference $\Delta E = E-2E_\gamma$. After this subtraction, the largest term in the resulting eigenvalue equation is given by the part proportional to $2\gamma/L$ in $H_{KP}$. This prefactor is multiplying a matrix whose elements are all equal to~1. Such matrix has one very large positive eigenvalue and a zero eigenvalue whose eigenvectors satisfy:
%%%%%%%%%%%%%%%%%%%
%%%%%%%%%%%%%%%%%%%
\begin{equation}
 \sum_K c_K = 0 \,.
 \label{eq:sum_coeffs}
\end{equation}
%%%%%%%%%%%%%%%%%%%
%%%%%%%%%%%%%%%%%%%
\noindent Because we are interested in the approximate ground state, the coefficients $c_K$ must obey this condition (see details in Appendix \ref{sec:analytical}).

Imposing Eq.~\eqref{eq:sum_coeffs} on the set of linear equations we obtain the condition on the energy:
%%%%%%%%%%%%%%%%%%%
%%%%%%%%%%%%%%%%%%%
\begin{equation}
 \sum_{K>0} \left(\frac{\hbar^2 K^2}{2m\Delta E}-1\right)^{-1} = \frac{1}{2} \,,
 \label{eq:energy_condition}
\end{equation}
%%%%%%%%%%%%%%%%%%%
%%%%%%%%%%%%%%%%%%%
\noindent which is satisfied for
%%%%%%%%%%%%%%%%%%%
%%%%%%%%%%%%%%%%%%%
\begin{equation}
 \Delta E = \frac{1}{2m} \frac{\hbar^2 \pi^2}{L^2} \,,
\end{equation}
%%%%%%%%%%%%%%%%%%%
%%%%%%%%%%%%%%%%%%%
\noindent in coincidence with the result of Eq.~\eqref{eq:energy_known} for $N=2$ to zero order in~$\epsilon$. And replacing the value of the energy we have that:
%%%%%%%%%%%%%%%%%%%
%%%%%%%%%%%%%%%%%%%
\begin{equation}
 c_K \simeq \frac{-2c_0}{4k^2-1}, \quad K = \frac{2\pi k}{L},\, k=1,2,\ldots
 \label{eq:solution}
\end{equation}
%%%%%%%%%%%%%%%%%%%
%%%%%%%%%%%%%%%%%%%
\noindent These coefficients correspond to the expansion of the function $\sin(\pi |x|/L)$ on the functions $\cos(2\pi k x/L)$. 

Plugging Eq.~\eqref{eq:solution} into $P(x|0)$ in the appropriate limit we consistently find that:
%%%%%%%%%%%%%%%%%%%
%%%%%%%%%%%%%%%%%%%
\begin{equation}
 P(x|0) \simeq \frac{2}{L} \sin^2(\pi x/L) .
 \label{eq:P_exact}
\end{equation}
%%%%%%%%%%%%%%%%%%%
%%%%%%%%%%%%%%%%%%%
\noindent The ground state for very strong attraction can then be understood in terms of two strongly bound pairs behaving as hard-core bosons. The function $\sin(\pi |x|/L)$ for the position of one pair relative to the other gives the necessary vanishing of the wave function for overlapping identical fermions while keeping the kinetic energy to a minimum. 

From Eq.~\eqref{eq:solution} one can also find the fidelity between the coboson ansatz and the correct ground state for $\lambda L\to \infty$:
%%%%%%%%%%%%%%%%%%%
%%%%%%%%%%%%%%%%%%%
\begin{equation}
 \mathcal{F} \simeq \left(1+2 \sum_{j=1}^{\infty} \frac{1}{(4j^2-1)^2} \right)^{-1} = \frac{8}{\pi^2} \simeq 0.811 ,
\end{equation}
%%%%%%%%%%%%%%%%%%%
%%%%%%%%%%%%%%%%%%%
\noindent which coincides with the results of the discrete model studied in Ref.~\cite{Cespedes_2019}. There is also a simple explanation for this value: it is the squared overlap between the normalized constant wavefunction and $\sqrt{2}\sin(\pi |x| /L)$ for $x\in[-L/2,L/2)$.

The previous results are of order zero in $\epsilon$ but directly neglecting the first order in $H$ and $S$ leads to wrong results. On the contrary, the steps above, which are detailed in Appendix \ref{sec:analytical}, recover the right behavior. Notice also that the  procedure is approximate and involves a cutoff. Equation~(\ref{eq:solution}) is only valid in the low-$K$ range, which in turn relies on the condition $\lambda L \gg 1$. 

\subsection{Numerical results for very strong attraction}

More accurate results for small but non-zero $\epsilon$ can be obtained from the numerical solution of the equations. We thus look for the approximate ground state from the generalized eigenvalue problem $H c = E S c$, where $S$ and $H$ are the matrices defined in Eqs.~(\ref{eq:Sdef}) and (\ref{eq:Hmatrix}) respectively. One should be careful that the number of states included is such that neglecting the basis states for which the relative motion is excited is justified. For $\lambda L \gg 1$, this still allows one to take many values of $K$ while keeping an energy difference with the states left out of the description. Alternatively, one can also include higher states for the relative motion; such a procedure is a generalization of the one presented above.

\begin{figure}
 \centering\includegraphics[width=0.9\columnwidth]{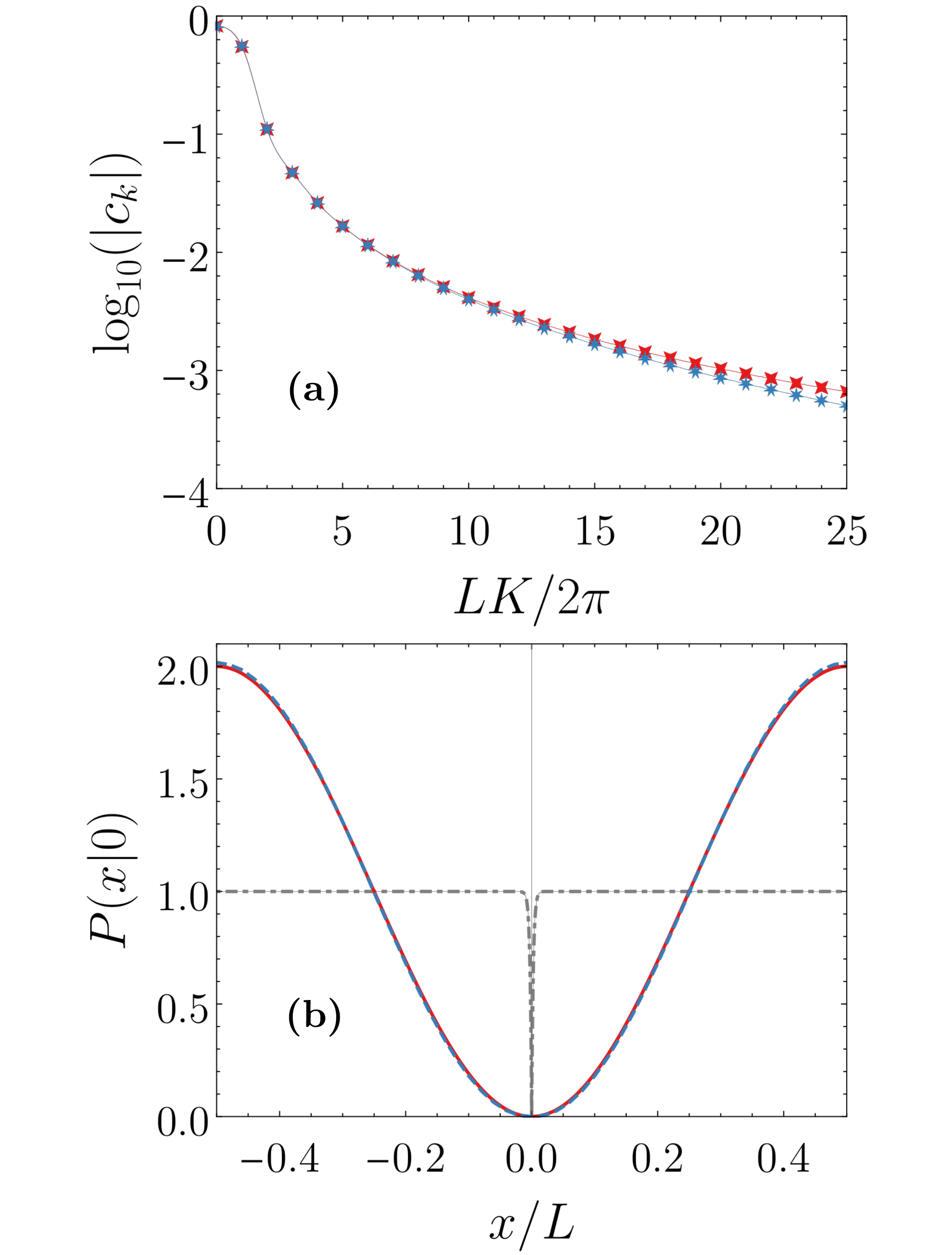}
 \caption{(a) Logarithm of the absolute values of the coefficients obtained from the numerical calculation for $\lambda L =200$ (blue seven-pointed stars) and from the analytic expression for infinite attractive interaction (red four-pointed stars). (b) Probability of finding a particle of kind $a$ at position $x$ given that another particle of the same kind was found at position 0 obtained from the numerical calculation for $\lambda L = 200$ (dashed blue line) and from the analytical solution (solid red line) in Eq.~\eqref{eq:P_exact}. The probability given by the coboson ansatz, see Eq.~\eqref{eq:prob_cob}, is shown in dot-dashed grey line. 
 \label{fig:ck_P}}
\end{figure}

As an example, we consider a case with $\lambda L = 200$ and 26 basis states. The obtained energy $\Delta E$ differs from the reference result in  Eq.~\eqref{eq:energy_known} by about 1.7\%, and the fidelity between the obtained ground state and the coboson ansatz is $\mathcal{F} \simeq 0.814$. In Fig.~\ref{fig:ck_P}~(a) we show the behavior of the coefficients (taking the logarithm of their absolute values) as a function of $KL/(2\pi)$. The agreement between the numerical solution (blue seven-pointed stars) and the analytical limit (red four-pointed stars) is very good for the largest coefficients, but the decay of the numerical solution is faster. As can be seen in Fig.~\ref{fig:ck_P}~(b)  the conditional probabilities calculated from the numerics (dashed blue line) and the analytical solution (solid red line) are also very similar, but the numerical solution takes a slightly larger value (by about 1\%) for opposing positions of the pairs. Figure~\ref{fig:ck_P}~(b) also depicts the probability in Eq.~ \eqref{eq:prob_cob}, calculated from the coboson ansatz (dot-dashed grey curve), which gives an abrupt decay of the probability around $x=0$ while the correct ground state shows a smooth behavior. 

%%%%%%%%%%%%%%%%%%%
%%%%%%%%%%%%%%%%%%%
%%%%%%%%%%%%%%%%%%%
%%%%%%%%%%%%%%%%%%%

%%%%%%%%%%%%%%%%%%%
%%%%%%%%%%%%%%%%%%%
% Sect: Concluding remarks.
%%%%%%%%%%%%%%%%%%%
%%%%%%%%%%%%%%%%%%%
\section{Concluding remarks}
\label{sec:conclusion}

Focusing on a simple 1D system of fermion pairs interacting through a delta potential we show that the coboson ansatz predicts a ground-state energy which is much larger than its exact value. The ansatz fails to provide a good approximation of the ground state for strong interaction because it cannot capture the spatial correlations between pairs. This failure contradicts the expectation that dilute systems with high entanglement between constituents and short-ranged interactions should possess a coboson-like ground state. 

For very strong attraction, fermion pairs approach the behaviour of hard-core bosons, which in 1D can be treated through fermionization \cite{Girardeau_1960}. One could then conjecture that the failure of the ansatz is a consequence of the fermionized character of the ground state. However, the results for the discrete model in \cite{Cespedes_2019} show that the ansatz also fails for a 2D system with a strong anisotropy for which the mapping between hard-core bosons and fermions does not apply. Thus, the phenomenon is more subtle and the full determination of the regime of validity of the ansatz is still an open problem.

Although we focused on the case of two pairs only, the structure of coboson theory is such that, when the ansatz is valid, the predictions for $N$ pairs are based on dominant contributions from the cases $N=1,\,2$ determining the coefficients in an expansion in powers of the density of particles~\cite{Combescot_2008}. Because of this, our 2-pair study shows that the ansatz does not provide a proper description of the ground state for arbitrary $N$. 

Our results represent a warning sign against the light use of the coboson ansatz in the description of problems such as ultracold atomic dimers in 1D traps. On the other hand, we also showed how the coboson formalism can be applied to the system even when the coboson ansatz fails. Therefore, the coboson formalism still constitutes a useful toolbox for strongly attractive interactions. In this regime, the treatment of the problem in the free-particle basis is highly costly while the coboson basis allows for straightforward calculations.

The problem we have tackled is solvable by means of the Bethe ansatz \cite{Yang_1967, Gaudin_1967}. Nevertheless, this procedure leads to cumbersome expressions for the wavefunction which are inconvenient for many calculations. On the contrary, our approach leads to expressions that are straightforward to manipulate. Although we have restricted to a very simple model, with only two pairs and exhibiting translational invariance, we are confident that similar strategies can be useful for broader families of physical systems.

%%%%%%%%%%%%%%%%%%%
%%%%%%%%%%%%%%%%%%%
\section*{Acknowledgments} 

The authors acknowledge grants from ANPCyT (Argentina) and MinCyT C\'ordoba (Argentina).

%%%%%%%%%%%%%%%%%%%
%%%%%%%%%%%%%%%%%%%
% Sect: Bib
%%%%%%%%%%%%%%%%%%%
%%%%%%%%%%%%%%%%%%%

\bibliographystyle{myieeetr}       
% Set the bibliography style to AMS
\bibliographystyle{h-physrev5}

%%%%%%%%%%%%%%%%%%%
%%%%%%%%%%%%%%%%%%%
\appendix
%%%%%%%%%%%%%%%%%%%
%%%%%%%%%%%%%%%%%%%

%%%%%%%%%%%%%%%%%%%
%%%%%%%%%%%%%%%%%%%
% Ground state of the relative motion
%%%%%%%%%%%%%%%%%%%
%%%%%%%%%%%%%%%%%%%
\section{Ground state of the relative motion} \label{sec:relative}

We consider two particles of mass $m$ in a ring of length $L$ with a delta-type attractive interaction, 
%%%%%%%%%%%%%%%%
%%%%%%%%%%%%%%%%
%\qe
\begin{align}
%%%%%%%
\label{eq_H_full}
%%%%%%% 
{\cal H} \! = \! -\frac{\hbar^2}{2m}\!\left(\frac{\partial^2}{\partial x_a^2}+\frac{\partial^2}{\partial x_b^2}\right) - \gamma \delta(x_a-x_b),
\end{align}
%%%%%%%%%%%%%%%%
%%%%%%%%%%%%%%%%
\noindent where $x_a$ and $x_b$ are the positions of the particles with a contact interaction of strength $\gamma > 0$. Introducing the center of mass $x_{{\scriptscriptstyle CM}} = (x_a + x_b)/2$ and relative coordinates $x_r = x_a-x_b$ the total Hamiltonian decouples ${\cal H}={\cal H}_{{\scriptscriptstyle CM}}+{\cal H}_r$, with ${\cal H}_{{\scriptscriptstyle CM}}$ and ${\cal H}_r$ describing the center of mass and relative wavefunction 
%%%%%%%%%%%%%%%%
%%%%%%%%%%%%%%%%
\begin{align}
%%%%%%%
\label{eq_HR_Hr}
%%%%%%% 
{\cal H}_{{\scriptscriptstyle CM}} & =  -\frac{\hbar^2}{2 m_{{\scriptscriptstyle CM}}}\frac{d^2}{dx_{{\scriptscriptstyle CM}}^2} , \\
{\cal H}_r & =  -\frac{\hbar^2}{2 m_r}\frac{d^2}{d x_r^2} - \gamma \delta(x_r), 
\end{align}
%%%%%%%%%%%%%%%%
%%%%%%%%%%%%%%%%
\noindent where $m_{{\scriptscriptstyle CM}} = 2m$ and $m_r = m/2$. The solutions are products of the form $\psi =  \psi_{{\scriptscriptstyle CM}} \, \psi_r$ with a total energy given by the sum of the relative and center of mass energies. The center of mass wave-functions are the free-particle states while the relative normalized wavefunctions are
%%%%%%%%%%%%%%%%
%%%%%%%%%%%%%%%%
\begin{equation}
%%%%%%%
\label{eq_psi_rel}
%%%%%%% 
\psi_r (x_r) = \frac{\cosh \left( \lambda (|x_r|-\frac{L}{2}) \right) }{\sqrt{\frac{L}{2} + \frac{\sinh (\lambda L) }{2 \lambda}}} ,
\end{equation}
%%%%%%%%%%%%%%%%
%%%%%%%%%%%%%%%%
\noindent with energy
%%%%%%%%%%%%%%%%
%%%%%%%%%%%%%%%%
\begin{equation}
%%%%%%%
\label{eq_energy_rel}
%%%%%%% 
E_{\gamma} = - \frac{\hbar^2 \lambda^2}{m} ,
\end{equation}
%%%%%%%%%%%%%%%%
%%%%%%%%%%%%%%%%
\noindent where $\lambda$ is determined by solving
%%%%%%%%%%%%%%%%
%%%%%%%%%%%%%%%%
\begin{equation}
%%%%%%%
\label{eq_lambda_rel}
%%%%%%% 
\left( \frac{\lambda L}{2} \right) \tanh \left( \frac{\lambda L}{2} \right) = \frac{\gamma L}{4} .
\end{equation}
%%%%%%%%%%%%%%%%
%%%%%%%%%%%%%%%%

\begin{figure}
 \centering\includegraphics[width=0.9\columnwidth]{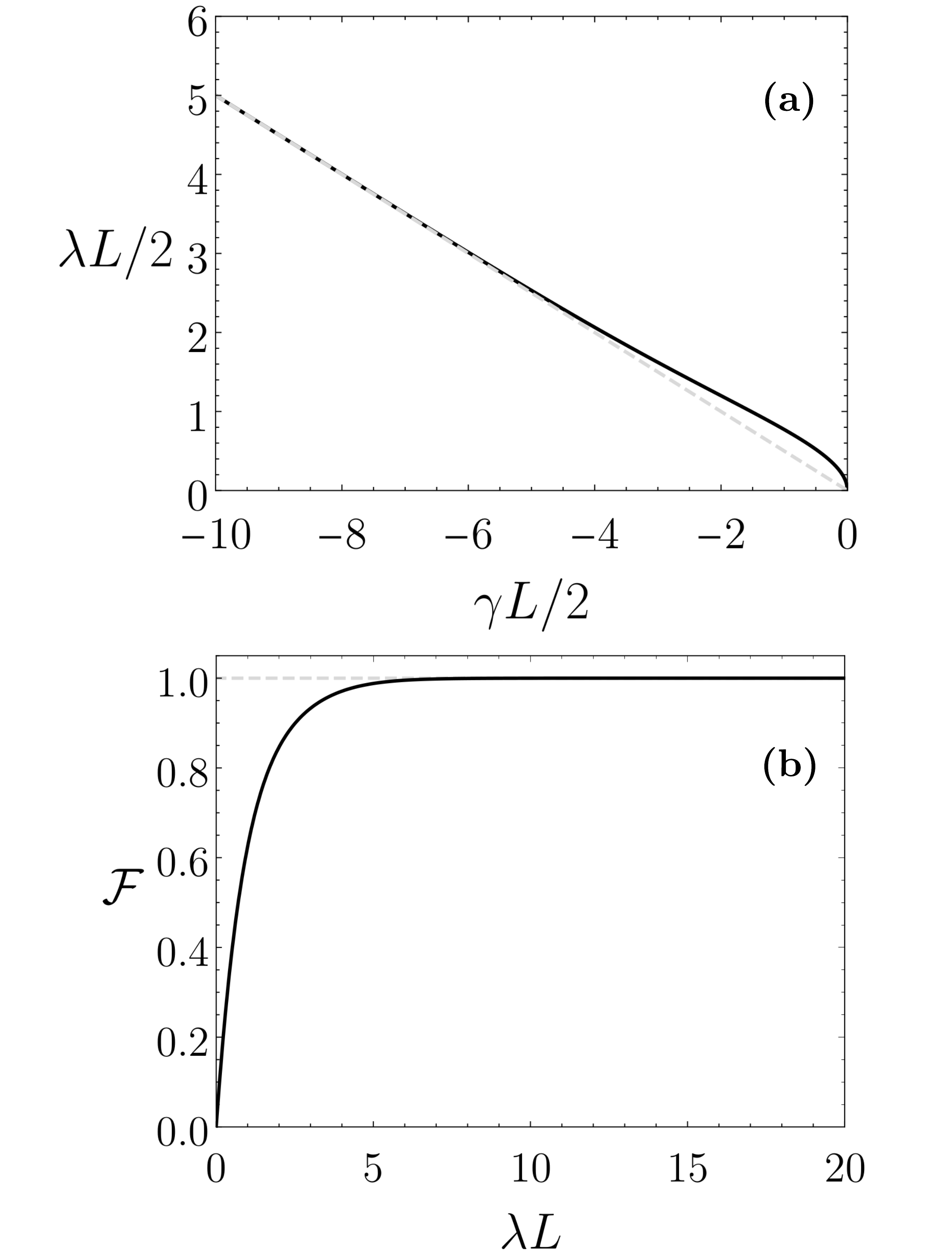}
 \caption{(a) Energy parameter $\lambda$ vs. the interaction strength $\gamma$ (black solid curve) and the corresponding approximation (see Eq.~\eqref{eq_lambda_rel_approx}) for $\gamma \gg 1$ (gray dashed line). (b)  Fidelity between the state and its approximation for $\gamma \gg 1$ as a function of the interaction strength.
 \label{fig_rel}}
\end{figure}

In the strong attractive regime $\gamma \gg 1$ the parameter $\lambda$ reads
%%%%%%%%%%%%%%%%
%%%%%%%%%%%%%%%%
\begin{equation}
%%%%%%%
\label{eq_lambda_rel_approx}
%%%%%%% 
\lambda \simeq \frac{m \gamma}{2 \hbar^2} ,
\end{equation}
%%%%%%%%%%%%%%%%
%%%%%%%%%%%%%%%%
\noindent while the relative wavefunction can be approximated by
%%%%%%%%%%%%%%%%
%%%%%%%%%%%%%%%%
\begin{equation}
%%%%%%%
\label{eq_psi_rel_approx}
%%%%%%% 
\psi_r (x_r) \simeq \sqrt{\lambda}\,\, e^{-\lambda |x_r|} .
\end{equation}
%%%%%%%%%%%%%%%%
%%%%%%%%%%%%%%%%
\noindent Figure \ref{fig_rel}~(a) depicts the $\lambda$ dependence on the interaction strength $\gamma$ (black solid curve) and the approximation of Eq.~\eqref{eq_lambda_rel_approx} (grey dashed line), showing a good agreement for $\gamma L/2 \gtrsim 5$. Panel~(b) gives the fidelity between the state and its approximation from Eq.~\eqref{eq_psi_rel_approx}, ${\cal F} = \vert \langle \psi_r (x_r)   \vert \psi_r^{\lambda L \gg 1} (x_r)\rangle  \vert^2$, which (as expected) goes to one for $\gamma L/2 \gtrsim 5$. 

%%%%%%%%%%%%%%%%%%%
%%%%%%%%%%%%%%%%%%%
%%%%%%%%%%%%%%%%%%%
%%%%%%%%%%%%%%%%%%%

%%%%%%%%%%%%%%%%%%%
%%%%%%%%%%%%%%%%%%%
% Schmidt decomposition of the single-pair ground state and associated criteria for bosonic behavior
%%%%%%%%%%%%%%%%%%%
%%%%%%%%%%%%%%%%%%%
\section{Schmidt decomposition of the single-pair ground state and associated criteria for bosonic behavior} \label{sec:schmidt}

The translational invariance of the ground state  for one pair makes the Schmidt decomposition straightforward. In fact, the state
%%%%%%%%%%%%%%%%
%%%%%%%%%%%%%%%%
\begin{equation}
%%%%%%%
\label{eq_psi_gs_twobody}
%%%%%%% 
\psi_{{\scriptscriptstyle GS}} (x_a,x_b) = \sqrt{\frac{\lambda}{L}}\,\, e^{-\lambda |x_a-x_b|} ,
\end{equation}
%%%%%%%%%%%%%%%%
%%%%%%%%%%%%%%%%
\noindent can be easily decomposed using:
%%%%%%%%%%%%%%%%%%%
%%%%%%%%%%%%%%%%%%%
\begin{align}
 \int_{-L/2}^{L/2} dx &\, e^{ikx-\lambda|x|} = 
 \nonumber \\
 & \frac{2\left\lbrace \lambda + e^{-\frac{\lambda L}{2}} \left( k \, \sin(\frac{k L}{2} ) - \lambda \, \cos(\frac{k L}{2}) \right) \right\rbrace}{\lambda^2+k^2} ,
\end{align}
%%%%%%%%%%%%%%%%%%%
%%%%%%%%%%%%%%%%%%%
\noindent that for $\lambda L \gg 1$ reduces to 
%%%%%%%%%%%%%%%%%%%
%%%%%%%%%%%%%%%%%%%
\begin{equation}
 \int_{-L/2}^{L/2} dx\, e^{ikx-\lambda|x|} \simeq \frac{2\lambda}{\lambda^2+k^2} ,
 \label{eq_integral}
\end{equation}
%%%%%%%%%%%%%%%%%%%
%%%%%%%%%%%%%%%%%%%
\noindent where we neglected exponentially small terms; notice that this result is equivalent to taking the limits of the integrals to infinity with an exponentially small error. The Schmidt decomposition gives
%%%%%%%%%%%%%%%%
%%%%%%%%%%%%%%%%
\begin{equation}
 \psi_{{\scriptscriptstyle GS}} (x_a,x_b)  = \sum\limits_{K} \frac{2}{\sqrt{\lambda L}} \left( \frac{1}{1+ \left( \frac{K}{\lambda}\right)^2} \right) \frac{e^{i K x_a}}{\sqrt{L}} \frac{e^{-i K x_b}}{\sqrt{L}} ,
\end{equation}
%%%%%%%%%%%%%%%%%%%
%%%%%%%%%%%%%%%%%%%
\noindent where we write the Schmidt modes or single particle orbitals as plane waves with $K = 2 \pi k/ L$, $k \in \mathbb{Z}$ (the decomposition is not unique and can also be given in terms of sine and cosine functions). With this procedure one can also calculate the Schmidt decomposition of the excited states.

The above expression allows the exact computation of entanglement measures, as well as several quantities used within the coboson formalism. We note that the calculation of the purity $\mathcal{P}$ of the reduced states of the constituent fermions does not require the Schmidt decomposition since it can be obtained from the normalization of the ground state as:
%%%%%%%%%%%%%%%%%%%
%%%%%%%%%%%%%%%%%%%
\begin{equation}
    \mathcal{P} = 1-\frac{S_{00}}{2} = \frac{5}{2\lambda L} \,.
\end{equation}
%%%%%%%%%%%%%%%%%%%
%%%%%%%%%%%%%%%%%%%
\noindent According to Refs. \cite{Law_2005, chudzicki_2010}, ideal bosonic behavior is expected when the purity is sufficiently small, or equivalently for high amounts of entanglement in the ground state.      

In Ref. \cite{ramanathan_2011} the authors propose a majorization criterion for good bosonic behavior arguing that in a case with no interactions condensation can be regarded as a deterministic LOCC (local quantum operation and classical communication) process. They take as initial state a product state of $N$ cobosons in $N$ different potential wells and as final state the $N$-particle coboson state $\ket{N}$. Then, they reason that a good bosonic behavior will be guaranteed when the initial and final state obey the majorization criterion for deterministic LOCC. Such criterion affirms that the vector of descending ordered eigenvalues of the initial state ($\vec{s}^{\,\,i} = \left\lbrace s_0^N, s_0^{N-1} s_1 \ldots \right\rbrace$) must be majorized by the ordered eigenvalues vector of the final state ($\vec{s}^{\,\,f} = \left\lbrace s_0 s_1 \ldots s_{N-1}, \ldots \right\rbrace/\tilde{\chi}_N$), i.e. $\vec{s}^{\,\,i} \prec \vec{s}^{\,\,f}$, where the $s_j$ denote the Schmidt coefficients of the two-body ground state decomposition and $\tilde{\chi}_N =  \sum_{j_1 < j_2 < \ldots < j_N} s_{j_1} s_{j_2} \ldots s_{j_N} < 1 $ is a normalization factor. 

In our problem, the ordered distribution of Schmidt coefficients of one pair takes the form:
%%%%%%%%%%%%%%%%%%%
%%%%%%%%%%%%%%%%%%%
\begin{equation}
    s_j = \frac{4}{\lambda L} \frac{1}{\left[1+\left(\frac{2\pi}{\lambda L}\right)^2 (\lceil \frac{j}{2} \rceil)^2\right]^2},
\end{equation}
%%%%%%%%%%%%%%%%%%%
%%%%%%%%%%%%%%%%%%%
\noindent where the symbol $\lceil x \rceil$ denotes the ceiling function. For simplicity we focus here on the case $N=2$, which was discussed in detail in the body of the paper. Writing the exact distributions for the initial and final states for the protocol proposed in \cite{ramanathan_2011} is complicated and beyond the scope of this study. However, it is important to notice that for $\lambda L\gg1$ our distribution of coefficients $s_j$ is very smooth and admits a continuous approximation. Furthermore, in the same limit the elements $s_j^2$ are very few in comparison with the rest, of the form $s_js_l$ with $j\neq l$. This means we can neglect their contribution. The Schmidt coefficients of the initial and final states are then essentially the same, except that each element $s_j s_l$ appears twice in the initial state (corresponding to the two possible choices picking from the two different wells) but only once in the final state. Thus, making a continuous approximation, the distributions of Schmidt coefficients for sufficiently large $\lambda L$ are related by:
%%%%%%%%%%%%%%%%%%%
%%%%%%%%%%%%%%%%%%%
\begin{equation}
    s^f(x)\simeq 2s^i(2x).
\end{equation}
%%%%%%%%%%%%%%%%%%%
%%%%%%%%%%%%%%%%%%%

\begin{figure}
 \centering\includegraphics[width=0.9\columnwidth]{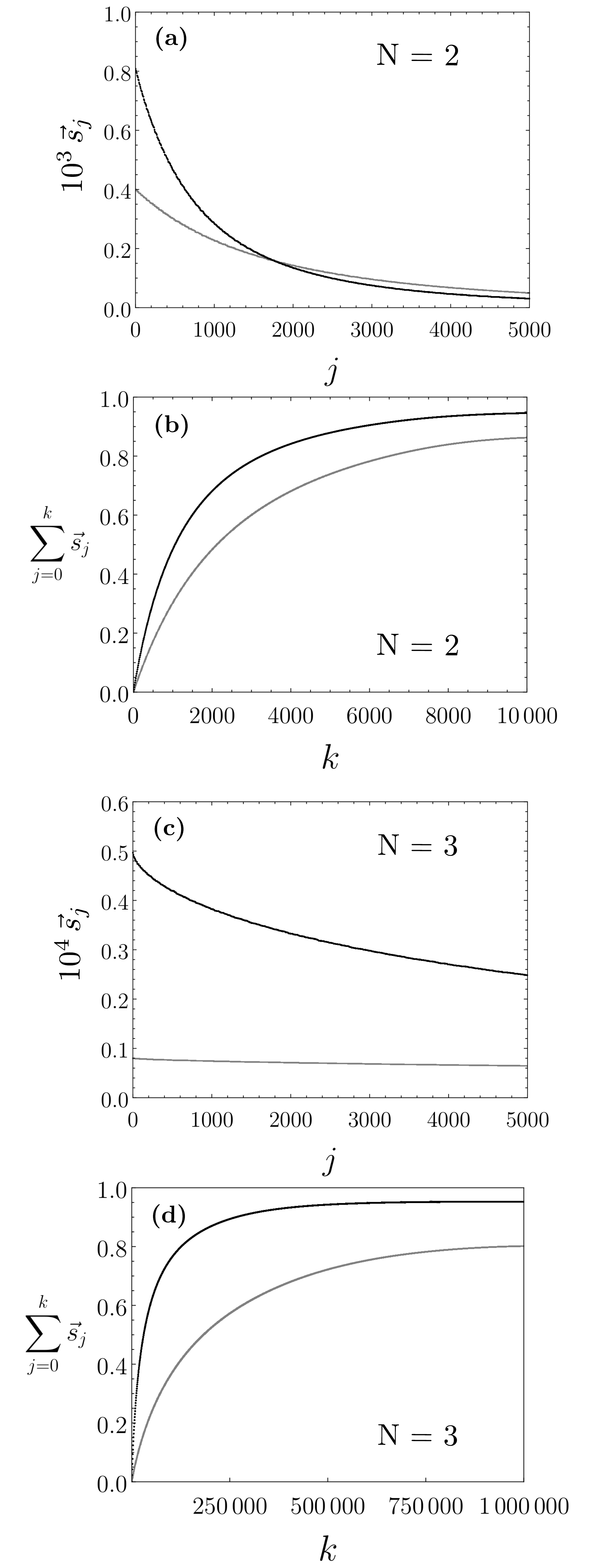}
 \caption{(a) Schmidt coefficients distribution $\vec{s}^{\,\,i}$ (grey squares) and $\vec{s}^{\,\,f}$ (black dots) for $\lambda L = 200$ and $N=2$. (b) Sum of the first $k$ ordered Schmidt coefficients for each of the distributions in (a) showing the numerical verification of the majorization criterion ($\vec{s}^{\,\,i} \prec \vec{s}^{\,\,f}$) up to $10^4$ coefficients for $N=2$ with the same color scheme as before. (c) Same as (a) for $N=3$. (d) Same as (b) for $N=3$ and $10^6$ coefficients. }
 \label{fig_maj}
\end{figure}

This shows that the majorization criterion is fulfilled and our ground state is a good bosonic state according to \cite{ramanathan_2011}. Indeed, we have:
%%%%%%%%%%%%%%%%%%%
%%%%%%%%%%%%%%%%%%%
\begin{equation}
    \int_0^M dx\, s^f(x) \simeq \int_0^{2M} dx\, s^i(x) > \int_0^M dx\, s^i(x) \,.
\end{equation}
%%%%%%%%%%%%%%%%%%%
%%%%%%%%%%%%%%%%%%%
\noindent This line of reasoning can also be extended to higher values of $N$, as long as $\lambda L$ is sufficiently large. As an example, in Fig.~\ref{fig_maj} (a) we show the Schmidt coefficients for initial (grey) and final (black) states for the case $N=2$ and $\lambda L = 200$ as in the figure of the main text. Figure \ref{fig_maj} (b) depicts the numerical corroboration of the majorization criterion $\vec{s}^{\,\,i} \prec \vec{s}^{\,\,f}$ up to $10^4$ coefficients for $N=2$. Panel (c) corresponds to $N=3$ and gives the same as panel (a) while panel (d) shows the numerical verification of the majorization criterion up to $10^6$ coefficients for $N=3$. 

%%%%%%%%%%%%%%%%%%%
%%%%%%%%%%%%%%%%%%%
%%%%%%%%%%%%%%%%%%%
%%%%%%%%%%%%%%%%%%%

Another necessary criterion for good bosonic behavior that has been  proposed states that $s_j \ll 1/N$. Such a criterion is based on the idea that a low density prevents the overlap between the wavefunction of the fermionic constituents. When ordering the Schmidt coefficients as $s_0 \geq s_1 \geq \ldots$ it is sufficient to ask for $s_0 \ll 1/N$, which in our case leads to $4 N \ll \lambda L$. Particularizing for $N=2$ it gives $8 \ll \lambda L$, condition satisfied throughout our derivations.

%%%%%%%%%%%%%%%%%%%
%%%%%%%%%%%%%%%%%%%
% Calculation of overlap and Hamiltonian matrices
%%%%%%%%%%%%%%%%%%%
%%%%%%%%%%%%%%%%%%%
\section{Calculation of the overlap and Hamiltonian matrices} \label{sec:calculations}

We start from the expression:
%%%%%%%%%%%%%%%%%%%
%%%%%%%%%%%%%%%%%%%
\begin{multline}
S_{PK} = [\delta_{KP} - \bra{P}\otimes\bra{-P} X_a \ket{K}\otimes\ket{-K} \,]\\ 
+ K\leftrightarrow -K \,,
\end{multline}
%%%%%%%%%%%%%%%%%%%
%%%%%%%%%%%%%%%%%%%
\noindent where we take $K, P \geq 0$. We note that this is a mere rewriting of the expressions in Ref. \cite{Combescot_2008} making use of an auxiliary operator $X_a$. This operator is defined in such a way that, when acting on two coboson states treated as made from distinguishable particles, it swaps the states of the fermions of kind $a$:
%%%%%%%%%%%%%%%%%%%
%%%%%%%%%%%%%%%%%%%
\begin{equation}
 X_a (\ket{k_a,k_b} \otimes \ket{q_a,q_b} ) = \ket{q_a,k_b} \otimes \ket{k_a,q_b} \,.
\end{equation}
%%%%%%%%%%%%%%%%%%%
%%%%%%%%%%%%%%%%%%%
\noindent In the language of coboson theory, this is a ``fermion exchange'' operation. For the case we consider the two terms with fermion exchanges are equal, so we may further reduce this to:
%%%%%%%%%%%%%%%%%%%
%%%%%%%%%%%%%%%%%%%
\begin{multline}
S_{PK} = \delta_{KP} (1 + \delta_{K,0}) \\
- 2 \bra{P}\otimes\bra{-P} X_a \ket{K}\otimes\ket{-K} \,.
\end{multline}
%%%%%%%%%%%%%%%%%%%
%%%%%%%%%%%%%%%%%%%

We now proceed with the overlap replacing the expressions for the coboson operators as in Eq.~(\ref{eq:coboson_field_basis}) of the main text. This gives a long expression that may be simplified by bringing all annihilation operators to the right and obtaining the corresponding delta functions from the anticommutators:  
%%%%%%%%%%%%%%%%%%%
%%%%%%%%%%%%%%%%%%%
\begin{equation}
 \{\Psi_\alpha(x), \Psi_\alpha^\dagger(x')\} = \delta(x-x') \,,
\end{equation}
%%%%%%%%%%%%%%%%%%%
%%%%%%%%%%%%%%%%%%%
\noindent with $\alpha=a,b$. The resulting delta functions eliminate several of the integrals and one finds:
%%%%%%%%%%%%%%%%%%%
%%%%%%%%%%%%%%%%%%%
\begin{align}
  \bra{P}\otimes\bra{-P} X_a & \ket{K}\otimes\ket{-K} = \frac{\lambda^2}{L}\int dx_1 \ldots dx_4 
  \nonumber \\
  & e^{ i \left\lbrace x_1 (P+K)/2+x_2(K-P)/2-x_3 K \right\rbrace }
  \nonumber \\
  & e^{-\lambda \sum\limits_j |x_j|} \,\delta(x_1+x_2-x_3-x_4) \,.
\end{align}
%%%%%%%%%%%%%%%%%%%
%%%%%%%%%%%%%%%%%%%

We solve these integrals by writing the remaining Dirac delta as:
%%%%%%%%%%%%%%%%%%%
%%%%%%%%%%%%%%%%%%%
\begin{equation}
 \delta(x) = \frac{1}{2\pi} \int du \,e^{iux} \,.
\end{equation}
%%%%%%%%%%%%%%%%%%%
%%%%%%%%%%%%%%%%%%%
\noindent Then, we arrive at an expression of the form:
%%%%%%%%%%%%%%%%%%%
%%%%%%%%%%%%%%%%%%%
\begin{align}
  \bra{P}\otimes\bra{-P} X_a \ket{K}\otimes\ket{-K} = & \frac{\lambda^2}{2 \pi L} \int du \int dx_1 \ldots dx_4 \nonumber \\
  & e^{i \sum\limits_j k_j x_j - \lambda \sum\limits_j |x_j| } \,,
\end{align}
%%%%%%%%%%%%%%%%%%%
%%%%%%%%%%%%%%%%%%%
\noindent where $k_1 = u+(P+K)/2$, $k_2=u+(K-P)/2$, $k_3=u-K$, $k_4=u$. Now we can use that the integrals in the $x_j$ factorize.  Then, using Eq.~\eqref{eq_integral} we are left with:
%%%%%%%%%%%%%%%%%%%
%%%%%%%%%%%%%%%%%%%
\begin{multline}
  \bra{P}\otimes\bra{-P} X_a \ket{K}\otimes\ket{-K} = \frac{\lambda^6}{L} \frac{8}{\pi} \\
  \int du \, \prod_j \frac{1}{\lambda^2+k_j(u)^2} \,.
\end{multline}
%%%%%%%%%%%%%%%%%%%
%%%%%%%%%%%%%%%%%%%
\noindent The remaining integral can be solved using residues and leads to Eq.~\eqref{eq:S} of the main text.

%%%%%%%%%%%%%%%%%%%
%%%%%%%%%%%%%%%%%%%
% Analytical approximation for the very strong coupling limit
%%%%%%%%%%%%%%%%%%%
%%%%%%%%%%%%%%%%%%%
\section{Analytical approximation for the very strong coupling limit}
\label{sec:analytical}

Here we provide the steps that lead to the analytical solution for the ground state for very large $\lambda L$. For this, we use the Taylor expansion in $\epsilon=1/(\lambda L)$ of the overlap and Hamiltonian matrices, taking up to first order in $\epsilon$. These expansions are provided in Eqs.~\eqref{eq:S_Taylor} and \eqref{eq:H_Taylor} of the main text respectively. As explained in the main text, we subtract the binding energy from $H$ and then look for the energy difference $\Delta E=E-2E_\gamma$.

The generalized eigenvalue equation $Hc=ESc$ then takes the form:
%%%%%%%%%%%%%%%%%%%
%%%%%%%%%%%%%%%%%%%
\begin{multline}
\sum_P \left[\frac{2\gamma}{L} + \delta_{KP} \frac{\hbar^2 P^2}{2m} - \frac{11 \epsilon \hbar^2}{4 m} (P^2+K^2) \right] c_P \\
= \Delta E \sum_P [\delta_{KP} (1+\delta_{K0}) -5\epsilon ] c_P\,.
\end{multline}
%%%%%%%%%%%%%%%%%%%
%%%%%%%%%%%%%%%%%%%
\noindent This can be rewritten as:
%%%%%%%%%%%%%%%%%%%
%%%%%%%%%%%%%%%%%%%
\begin{multline}
\left(\frac{2\gamma}{L} - \frac{11 \epsilon \hbar^2}{4 m} K^2\right) \sum_P c_P + \frac{\hbar^2 K^2}{2m} c_K - \sum_P \frac{11 \epsilon \hbar^2}{4 m} P^2 c_P\\
= \Delta E \left[ c_K (1+\delta_{K0}) -5\epsilon \sum_P c_P \right]\,.
\end{multline}
%%%%%%%%%%%%%%%%%%%
%%%%%%%%%%%%%%%%%%%
\noindent It is convenient to divide every term by a reference energy $E_1 = \hbar^2 K_1^2/(2m)$, with $K_1=2\pi/L$, so that we get:
%%%%%%%%%%%%%%%%%%%
%%%%%%%%%%%%%%%%%%%
\begin{multline}
\frac{1}{\epsilon} \left(\frac{2}{\pi^2} - \frac{11 \epsilon^2}{2} \frac{K^2}{K_1^2} \right) \sum_P c_P + \frac{K^2}{K_1^2} c_K - \frac{11 \epsilon}{2} \sum_P  \frac{P^2}{K_1^2} c_P\\
= \frac{\Delta E}{E_1} \left[ c_K (1+\delta_{K0}) -5\epsilon \sum_P c_P \right] \,.
\end{multline}
%%%%%%%%%%%%%%%%%%%
%%%%%%%%%%%%%%%%%%%
\noindent Assuming that we only include in our description values of $K$ so that $K/K_1$ is still of order 1, we see that we must have:
%%%%%%%%%%%%%%%%%%%
%%%%%%%%%%%%%%%%%%%
\begin{equation}
    \sum_P c_P = \mathcal{O}(\epsilon) \,.
    \label{eq:coef_sum_epsilon}
\end{equation}
%%%%%%%%%%%%%%%%%%%
%%%%%%%%%%%%%%%%%%%
\noindent In the next steps, we neglect all terms of order $\epsilon^2$ and after rearranging we obtain:
%%%%%%%%%%%%%%%%%%%
%%%%%%%%%%%%%%%%%%%
\begin{multline}
\frac{\Delta E}{E_1} c_K (1+\delta_{K0}) - \frac{K^2}{K_1^2} c_K \\
=\frac{1}{\epsilon} \frac{2}{\pi^2} \sum_P c_P  - \frac{11 \epsilon}{2} \sum_P  \frac{P^2}{K_1^2} c_P \,.
\end{multline}
%%%%%%%%%%%%%%%%%%%
%%%%%%%%%%%%%%%%%%%

The quantity on the right is of order 1 and does not depend on $K$. We can thus use that the left-hand side is independent of $K$ as well, and arrive at the condition:
%%%%%%%%%%%%%%%%%%%
%%%%%%%%%%%%%%%%%%%
\begin{equation}
    \left(\frac{\Delta E}{E_1} -  \frac{K^2}{K_1^2} \right) c_K = 2 \frac{\Delta E}{E_1} c_0,\quad K>0\,.
\end{equation}
%%%%%%%%%%%%%%%%%%%
%%%%%%%%%%%%%%%%%%%
\noindent Plugging this equality into condition~\eqref{eq:coef_sum_epsilon} for the sum of the coefficients, to zero order in $\epsilon$ we obtain Eq.~\eqref{eq:energy_condition} of the main text. %  

\end{document}